# Optimization of Location Management for PCS Networks with CTRW Mobility Model

Qinglin, Zhao, Soung C. Liew, *Senior Member*

*Abstract:* This paper considers the design of the optimal location-update area (LA) of the distance-based scheme for personal communication service (PCS) networks. We focus on the optimization of two design parameters associated with the LA: 1) initial position upon LA update; 2) distance threshold for triggering of LA update. Based on the popular continuous-time random walk (CTRW) mobility model, we propose a novel analytical framework that uses a diffusion equation to minimize the location management cost. In this framework, a number of measurable physical parameters, such as length of road section, angle between road sections, and road section crossing time, can be integrated into the system design. This framework allows us to easily evaluate the total cost under general call arrival distributions and LA of different shapes. For the particular case of circular LA and small Poisson call-arrival rate, we prove the following: (1) When the drift is weak, the optimal initial position approaches the center of the LA; when the drift is strong, it approaches the boundary of the LA. (2) Comparing the optimal initial-position and center-initial-position solutions (which is assumed in most prior work), when the drift is weak, the optimal distance threshold and the minimum total cost are roughly equal; when the drift is strong, the optimal distance threshold in the later is about $\sqrt[3]{2}$ ($\approx$1.260) times that in the former, and the minimum total cost in the later is about $\sqrt[3]{4}$ ($\approx$1.587) times that in the former. That is, optimizing on initial position, which previous work did not consider, has the potential of reducing the cost measure by 37%.

## I. INTRODUCTION

Personal communications services (PCS) networks provide wireless access to mobile users anywhere, anytime in an uninterrupted and seamless way, using advanced microcellular techniques [1]. For location management in PCS networks, the coverage area of PCS networks can be partitioned into a number of location-update areas (LAs). Each LA consists of a group of cells. The location of mobile terminals (MTs) is tracked based on LA. When an MT crosses the boundary of its current LA, a new LA (which partially overlaps with the old one), will be defined (see Section III and Fig. 2 for elaboration and illustration). The LA of the MT is stored in a network database. Upon a request to identify the location of the MT, the network pages all cells within its LA. Location management cost includes (i) the terminal paging cost, and (ii) the LA update cost. There is generally a trade-off between the two costs that depend on the number of cells per LA.

Many location management schemes have been proposed to minimize the location management cost. Among them, three basic dynamic schemes are frequently referenced: movement-based scheme [2], time-based scheme [3], and distance-based scheme [4]. In the above three schemes, an MT performs a location update whenever the number of crossed cell borders, the elapsed time, and the traveled distance, respectively, exceeds a predefined threshold. It has been proved in [4] that the distance-based scheme achieves better performance compared with the other two schemes. Our work here focuses on the distance-based scheme.

With the popular CTRW mobility model [9], which is widely used in physics, engineering, and related disciplines, this paper proposes a novel analytical framework based on a diffusion equation to minimize the total cost of the distance-based scheme. In this framework, the dependency of the update cost on the initial position of the MT is captured through a backward differential equation; and the dependency of the paging cost on the current position of the MT is captured through a forward differential equation. A salient feature of this framework is that it can be used to evaluate the cost impact of various physically measurable parameters, including the initial position, the LA shape, and the general call arrival distribution.

This paper derives analytical results that show how LA should be designed and how it is related to various physical parameters that characterize the movement of MT. Specifically, for the case of a circular LA, when the call arrival rate is small and there is a strong preference for an MT to move in a particular direction, we show that the initial position of the MT should be "off-centered" within the newly defined LA upon an LA update. Our analytical results quantify this optimal offset and indicate that as much as a 37% cost reduction can be achieved, regardless of the relative magnitudes of the underlying two constituent costs, terminal paging and LA update costs.

### A. Related Work

In most previous work, a widely used mobility model is the classical "discrete" random walk model [2], [4], [7]-[8]. This classical model assumes that the coverage of PCS networks is partitioned into many square or hexagon cells with same size. A movement from one cell to another is defined as one displacement. The displacements are only between adjacent cells (as opposed to roads of varying lengths), limiting the number of displacement directions to four or six. Furthermore, the displacement length and displacement time are constant. This model oversimplifies the movement characteristics of an MT and fails to capture many practical and fundamental movement characteristics.

Ref. [12] considered a special case of the more general CTRW model in this paper. As with this paper, [12] does not require the assumption of discrete movement across square or hexagonal cells. However, the "initial position" within an LA is limited to the origin; and the displacement time and displacement length are constant. Furthermore, the update cost due to call arrival is neglected. By contrast, this paper provides a more comprehensive and general treatment which culminates in the mean location update interval theorem of the location management (see Theorem 2).

There has been some previous work on the initial position and the LA shape. Ref. [13] investigated fractional Brownian motion with an initial position. However, the focus was on handoff management rather than location management. Location management addresses how to track an MT and how to reduce the tracking cost, whereas handoff management addresses how to maintain an ongoing communication and how to reduce the handoff delay and the packet loss. Ref. [14] considered adaptive LA with the initial position taken into account. However, the treatment was on one-dimensional motion, and update cost incurred by call arrivals is ignored. Ref [18] considered adaptive location update and irregular LA. Its analysis was based largely on the cell residency time distribution, while our paper here investigates the effect of more fine-grained information, e.g., layout and distribution of road sections within a cell (defined in Section II.C). Relative to the above work, a distinguishing feature of our treatment is that we attempt to make a connection between measurable mobility-related physical parameters and the design of LAs.

This rest of this paper is organized as follows. In Section II, we describe the model assumptions. Section III proposes the location management scheme based on the optimal initial position and the optimal distance threshold. Section IV proposes a general analytical framework to calculate the optimal initial position and the optimal distance threshold. Section V presents the numerical results to evaluate the location management cost. Section VI considers the implementation issue. Section VII concludes this paper.

## II. MODEL ASSUMPTIONS

This section presents the model assumptions used in this paper for the per-user distance-based scheme. With the exception of A2 and A7, these assumptions are widely adopted by the previous work [6], [10]-[13], [15]-[17], [19]. Further extensions for some assumptions can be found in Section IV.D.

### A. System model

The coverage area of PCS networks is a two-dimension plane that consists of many cells. Each cell has a base station. Each base station periodically broadcasts its physical coordinate as the identification (ID) of its cell. The networks and the MTs are connected by radio links between the base stations and the MTs. The network can track each MT through the LA, which consists of some cells and is dynamically formed based on each MT's current movement pattern. When a request to identify the location of an MT arrives, the network can page the MT within its current LA with certainty. For further analyses, we specify the following assumptions:

**A1**: The LA is a circular disc $\Omega$ with radius R.

**A2**: Each LA (rather than each cell) contains a number of interconnected road sections. Let $\xi$ denote the length of the road section. We require that $R \gg E(\xi)$, where $E(\cdot)$ denotes the mean of the $\cdot$. In other words, an MT typically travels a number of road sections before it moves out of its LA. $R \gg E(\xi)$ is a very strong condition. In fact, a moderate size of R is enough compared with $E(\xi)$. We will discuss this assumption in Section IV. D. (V).

### B. MT model

The MT can automatically call the system and identify itself and its current cell. It can also estimate its movement parameters, such as the mean displacement length, the mean displacement direction, and the mean displacement time. When a location update is performed, the MT will transmit these parameter values to the network so that the network can calculate the optimal LA based on these values.

### C. Mobility Model

In this paper, the popular CTRW mobility model is considered. We use this model to capture the movement characteristics of an MT, as follows. Assume that the coverage of PCS networks is an "infinite plane" on which there are many roads crisscrossing each other. The movements of the MT are governed by the layout of roads and paths. The MT may move along a road on a specific direction until it comes to a junction point. At the junction point, the MT may take a new direction. The process is repeated indefinitely in the model. We refer to the part of a road between two junction points as a road section and refer to the movement of the MT over a road section as one displacement. Thus, the displacement length corresponds to the length of a road section, the displacement direction corresponds to the angle between the direction of the selected road sections and a preferred direction explained later, and the displacement time corresponds to the time that the MT spends on the road section. The CTRW model makes the following assumptions. (see Fig. 3).

**A3**: All displacement lengths are i.i.d. with a random variable $\xi$ with a finite mean and a finite variance.

**A4**: All displacement directions are i.i.d. with a random variable $\Theta$ with a finite mean and a finite variance.

**A5**: All displacement time are i.i.d. with a random variable $\eta$ with a finite mean and a finite variance.

**A6**: Displacement length, displacement direction, and displacement time are independent each other.

Let $X_i$ be the i-th displacement vector described by ($\xi \cos \Theta$, $\xi \sin \Theta$). Let $T_i$ be the i-th displacement time described by $\eta$. From A3-A6, $\{(X_i, T_i), i \geq 1\}$ forms an i.i.d. sequence, where $X_i$ is independent of $T_i$. $\{(X_i, T_i), i\geq 1\}$ completely characterizes the movement of an MT. Mathematically, such movement process is called the CTRW. Here, we consider the simple version of the CTRW model in which the MT either remains fixed in an end point of the road section or makes an instantaneous displacement. When $X_i$ is a unit-length displacement limiting the number of displacement directions with equal angle to four or six; and $T_i$ is a unit time, the CTRW Model becomes the classical "discrete" random walk model. It has been pointed out in [9]: Although we regard the displacements as taking place instantaneously, with the MT resting for random times $T_i$, we can well interpret the MT as continually in motion, with the time $T_i$ taken from a road section to another.

Further, the following Theorem 1 states that $\{(X_i, T_i), i\geq 1\}$ can be characterized by a diffusion process $\{Y_t, t\geq 0\}$ related to the mean and the variance of $(\xi\cos\Theta, \xi\sin\Theta)$ and $\eta$, where $Y_t$ denotes the MT's position at time t. $Y_0$ denotes the MT's initial position, where $Y_0 = X \in \Omega$.

**Theorem 1** [9]**:** Under the assumptions A1-A6, if an MT performs lots of displacements before it moves out of the LA, then the MT's movement process $\{(X_i, T_i), i\geq 1\}$ can be approximated by a homogeneous diffusion process $\{Y_t, t\geq 0\}$ with a drift vector $\mu$ and a diffusion coefficient matrix $\sigma$, where $\mu$ and $\sigma$ are respectively given by

$$\mu \triangleq (\mu_1, \mu_2) = \frac{E(\bar{\xi})}{E(\eta)}, \ \bar{\xi} = (\xi\cos\Theta, \xi\sin\Theta) \quad (1)$$

$$\sigma \triangleq \begin{pmatrix} \sigma_{11} & \sigma_{12} \\ \sigma_{21} & \sigma_{22} \end{pmatrix} = \frac{Var(\bar{\xi})E^2(\eta) + Var(\eta)E(\bar{\xi})E(\bar{\xi})^T}{E^3(\eta)}$$

In (1), $E(\cdot)$, $Var(\cdot)$, and $(\cdot)^T$ denote the mean, the variance, and the transposition of the $\cdot$, respectively.

**Remarks:** The key point of the proof is to compute the Fourier-Laplace transform of the probability density of the MT's position at time t. Some simulations have shown that the diffusion process $\{Y_t, t\geq 0\}$ can well approach the original process $\{(X_i, T_i), i\geq 1\}$ [9]. Due to Theorem 1, hereafter, we use the diffusion process $\{Y_t, t\geq 0\}$ with parameters $\mu$, $\sigma$, and $Y_0 = X \in \Omega$ to describe the original movement $\{(X_i, T_i), i\geq 1\}$ with assumptions A1 to A6. Based on Theorem 1, this paper uses the differential equation (rather than the Fourier-Laplace transform method) to solve the optimization problem of location management.

Now, we set up the system coordinates as follows (see Fig. 3): The x-axis positive direction is the preferred direction. The y-axis direction is the upward direction that is orthogonal to the x-axis. The origin is the center of the LA. The initial position of the MT is always at the x-axis. This paper especially considers the location management when the displacement direction is preferred. Therefore, we make the further assumption as follows.

**A7**: The probability density function of $\Theta$, $f_\Theta(k, \theta)$, is symmetric with respect to the preferred direction (or the x-axis positive direction) and has the following properties:

$$f_\Theta(k, \theta) = f_\Theta(k, -\theta), \quad (2)$$

$$f_\Theta(k,\theta) = \begin{cases} 1/(2\pi), & k \to 0 \\ \delta(\theta), & k \to \infty \end{cases}$$

where $\delta(\cdot)$ is the one-dimensional delta function.

The (2) reflects the preference of an MT for a particular direction in its motion. The preferred direction is preserved throughout the motion of the MT even though it may momentarily travel at angle to the preferred direction because of the geographical layout of roads. The k in (2) is a factor which corresponds to the propensity for deviation of the direction of the new selected road away from the preferred direction. In the limit $k = 0$, an MT changes its direction of travel at an angle between $0^0$ and $360^0$ with uniform probability with respect to the preferred direction. In the limit $k = \infty$, the MT moves along the preferred direction with probability 1. The (3) below gives an example of $f_\Theta(k, \theta)$ and Fig. 1 plots the corresponding probability density function.

$$f_\Theta(k, \theta) = \frac{ke^{-k|\theta|}}{2(1-e^{-k\pi})} \quad (3)$$

$$0 \leq k \leq \infty, \ -\pi \leq \theta \leq \pi,$$

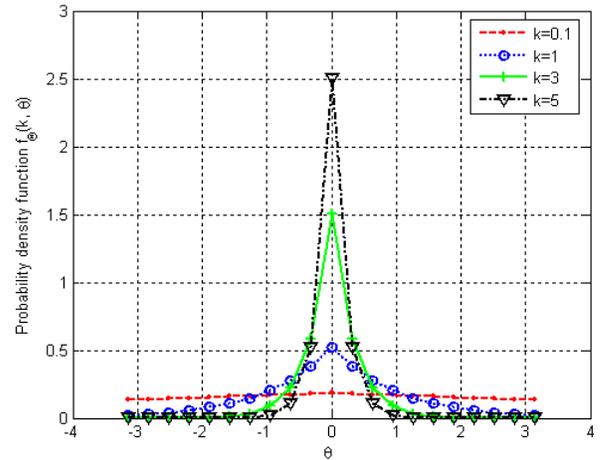

Fig. 1. The probability density function $f_\Theta(k, \theta)$.

### III. PROPOSED LOCATION MANAGEMENT STRATEGY

This paper focuses on the distance-based location management scheme. Location management scheme consists of two components. One is the location update strategy which defines the method that the MT notifies the network of its position. The other is the terminal paging strategy which defines the method that the network finds the called MT. For the rest of this paper, we use the term "network" to refer to the system that maintains the location database and that interacts with the MT to support location management in the network. Each entry in the location database is indexed by the identifier (ID) of an MT, and contains the information of the MT itself and the IDs of the cells within its current LA. Further details can be found in Section VI.

With the consideration of the preferred direction, the proposed distance-based scheme is different from the usual distance-based scheme in two aspects. First, in the usual scheme, the MT's initial position $Y_0$, where it has performed the last location update is always the center O of the LA. In contrast, in our scheme, the MT's initial position might be different from the center O of the LA. Second, when considering the paging strategy with delay constraint, in the usual scheme, the sub-paging areas are always symmetric with respect to the center O. In contrast, in our scheme, the sub-paging areas are always symmetric with respect to the preferred direction.

*A.    The usual distance-based scheme*

Fig. 2 illustrates the operation of the usual distance-based scheme. Location update operation: Assume that the MT's initial position is $Y_0$ and its current distance threshold is R. Let the $Y_0$ be the origin O. Then the origin O and the R define the current LA (called the old LA). When an MT moves within its associated old LA, location updates do not occur. However, when the MT moves out of the old LA, say at $Y_\tau$, a location update will be triggered (either by the MT or the network) and a new LA will be formed likewise. Then by the ID of an MT, the network can update the LA of the MT. As shown in Fig. 2, the new and the old LAs can partially overlap.

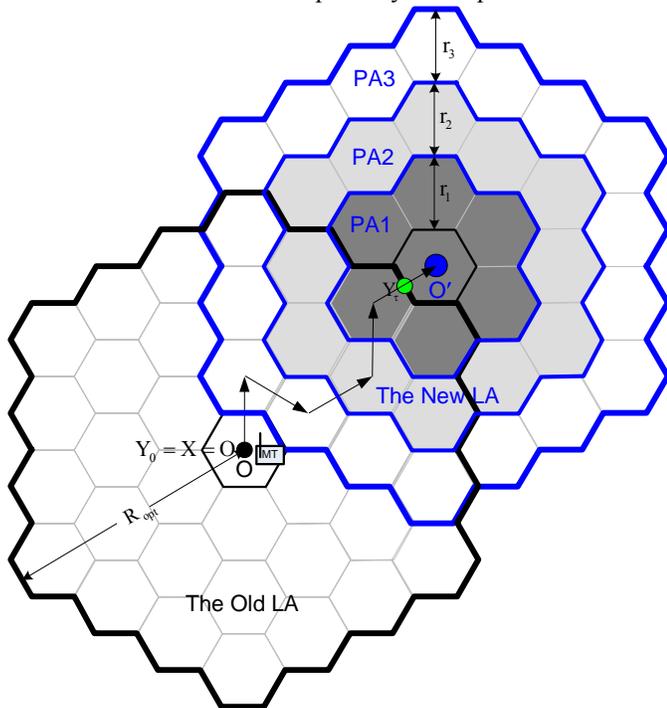

Fig. 2. The usual distance-based scheme.

Terminal paging operation: Assume the called MT is currently at the new LA. When a request calls the called MT, by the ID of the called MT, the network can retrieve the current LA of the MT. Then the network can find the MT through sending a paging message to all the cells of the current LA of the MT. Now we consider the paging scheme with delay constraint. Assume that the maximum paging delays is *m*. To reduce the paging cost, the network can partition the whole current LA into *m* sub-paging areas PA1, PA2, …, and PA*m* which are symmetric with respect to the center O. In the usual partition, these sub-paging areas are the concentric rings with equal-width. We call the difference between the outer radius and the inner radius of a ring the width of the ring. Fig. 2 shows 3 sub-paging areas: PA1 (in black), PA2 (in gray), and PA3 (in white), where $r_i$ is the width of the PAi. Note that PA1 includes the cell O to avoid triviality. Then the network can poll the sub-paging areas PA1, PA2, …, and PA*m* sequentially until the called MT is found.

In addition, during the paging operation, upon identifying the location and the cell the MT is currently residing at, a new LA will be formed. Therefore, a location update operation will also be triggered, which is often ignored by previous schemes. Also, the overlapping between the new LA and the old LA can reduce the "LA boundary crossing probability" in the future.

*B.    The proposed distance-based scheme*

Fig. 3 illustrates the operation of the proposed distance-based scheme. In the coordinate system in Fig. 3, the x-axis positive direction is always the preferred direction. The initial position of the MT X is always at the x-axis. The LA $\Omega$ is a circular disc with radius R. The origin O is the center of the LA. Location update operation: The location update operation in the proposed scheme is same to the one of the usual distance-based scheme except the design of the LA.

There are two key design issues in our proposed scheme: 1) How large should the LA be (i.e., in the circular LA model, what is the optimal radius R value)? 2) What should be the optimal offset when a new LA is formed (i.e., in the circular LA model, where is the center of the LA, or what is the optimal initial position X within an LA)? Note that the center will be determined once the initial position is determined. In contrast, the usual distance-based scheme only considers how to design the size of the LA since it regards the initial position as the center of the LA.

**The "optimal" LA size $R_{opt}$:** Since LA is a circular disc, the size of the LA is defined by its radius R. This radius R is just the well-known distance threshold in the location management scheme. Obviously, larger R will reduce the frequency of location update triggered by boundary crossing, hence the overall location-update cost. However, larger R also means that there are more cells in each LA so that the paging cost will be higher. There is an optimal distance threshold $R_{opt}$ to minimize the both costs given a characterization of the MT movement.

**The "optimal" initial position $X_{opt}$:** The location where the MT performs its last location update at the boundary of the old LA is the initial position of the MT in the new LA. In the usual design of LA, the center of the LA, O, is the MT's initial position X. In general, this design limitation is unnecessary. In this paper, the MT's movement is characterized by a diffusion process by Theorem 1. Intuitively, to reduce the frequency of location update (and hence reduce the location update cost), when the movement is an unbiased diffusion process, the center of the LA is indeed the initial position. However, when the movement is a biased diffusion process with a positive preferred direction, the MT's initial position that is at the negative x-axis can delay the exit time of the MT. Therefore, there is an optimal initial-position offset $X_{opt}$ with respect to O. The optimization problem on the initial position is to find the $X_{opt}$ so as to maximize the time before the next location update is triggered by boundary crossing given the $R_{opt}$.

Once we determine the optimal R and X, we can form a LA easily (further details on the construction of the LA can be found in the Section VI). Thereafter, the location update operation can completely follow that of the usual distance-based scheme. The following section will explore the expressions of $X_{opt}$ and $R_{opt}$.

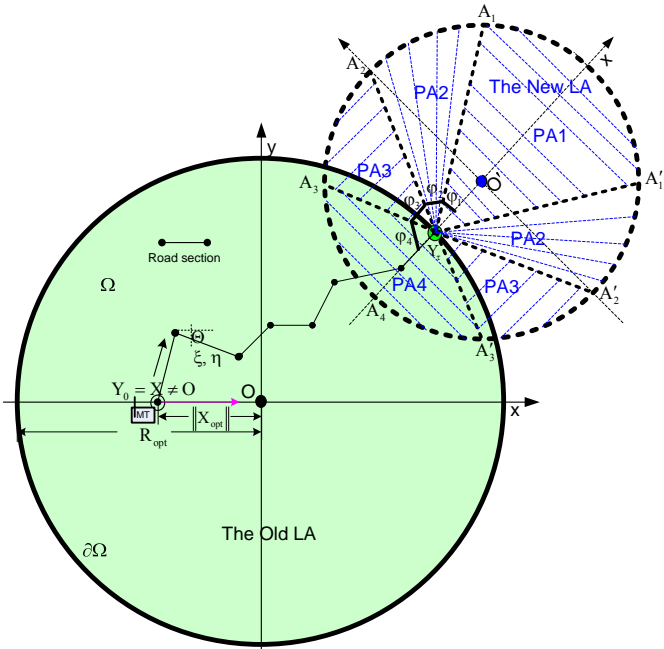

Fig. 3. The proposed distance-based scheme.

Terminal paging operation: The terminal paging operation in the proposed scheme is same to the one of the well-known distance-based scheme except the partition method of the paging area with delay constraint.

Assume that the maximum paging delays is m. The sub-paging areas PA1, PA2, …, and PAm in our scheme are symmetric with respect to the x-axis. Fig. 3 shows 4 sub-paging areas: PA1 (the region enclosed by $Y_\tau A_1 A_1'$), PA2 (the regions enclosed by $Y_\tau A_1 A_2$ and $Y_\tau A_1' A_2'$), PA3 (the regions enclosed by $Y_\tau A_2 A_3$ and $Y_\tau A_2' A_3'$), and PA4 (the region enclosed by $Y_\tau A_3 A_4 A_3'$).

Let $\varphi_1 = \angle A_1 Y_\tau O$, $\varphi_2 = \angle A_2 Y_\tau A_1$, $\varphi_3 = \angle A_3 Y_\tau A_2$, and $\varphi_4 = \angle A_4 Y_\tau A_3$, and so on. We call $\varphi_i$ is the angle of PAi. One method to assign the value of these angles is given by

$$\varphi_i = \begin{cases} \min(\pi, \text{Var}(\Theta)), & i = 1 \\ \dfrac{\pi - \min(\pi, \text{Var}(\Theta))}{m-1}, & 2 \leq i \leq m \end{cases} \quad (4)$$

where Var ($\Theta$) denotes the variance of the displacement direction $\Theta$.

Once we determine the sub-paging areas, we can easily differentiate different sub-paging areas by the cell coordinates (further details on the partition of the paging area can be found in the Section VI).

## IV. COST EVALUATION OF THE PROPOSED LOCATION MANAGEMENT STRATEGY

As explained in Section III. B, the location management cost are related to the initial position and the radius of the LA. This section optimizes the initial position and the radius so as to minimize the total cost that includes the location update cost and the terminal paging cost.

Location update cost $C_u$: The location update cost is related to the mean location update interval. In the location management strategy, when an MT moves out of its LA $\Omega$ or when a new call arrives at the MT, one location update operation will be triggered. Therefore, the mean location update interval is the mean of the minimum between the call arrival interval and the first exit time of the MT from LA defined as follows.

Let $\{Y_t, t\geq 0\}$ denote the MT's movement process stated in Theorem 1 with initial position $Y_0 = X \in \Omega$. The MT's first exit time $\tau (X, R)$ from $\Omega$ is given by

$$\tau (X, R) = \inf\{t \geq 0: Y_t \notin \Omega\} \quad (5)$$

Let $\zeta$ denote the call arrival interval with parameter $\lambda$. Then the mean location update interval $T(X, R, \lambda)$ can be defined as follows

$$T (X, R, \lambda) = E(\min\{\zeta, \tau (X, R)\}) \quad (6)$$

Let U denote the cost of performing a location update. Then, the mean location update cost per unit time is given by

$$C_u (X, R, \lambda) = U / T(X, R, \lambda) \quad (7)$$

Terminal paging cost $C_p$: We now consider the terminal paging cost with delay constraint. Assume that the maximum paging delay is m and that the sub-paging areas are PA1, PA2, …, and PAm. Let $P_i$ denote the probability that the called MT is within PAi when a call arrives at it. Let $A_i$ denote the area of PAi. Let V denote the cost of paging a cell. Assume that the area per cell is equal to 1. Then the mean terminal paging cost per unit time is given by

$$C_p(\lambda, R) = \lambda V \sum_{i=1}^{m} P_i A_i \quad (8)$$

$$P_i = \iint_{PA_i} p(Y, T) dY, \quad A_i = \iint_{PA_i} dY$$

where $T = T(X, R, \lambda)$ in (6) and $p(Y, t)$ is the conditional probability density function that MT's location is Y at time t given the initial location X. For the above process $\{Y_t, t \geq 0\}$, from the diffusion process theory, $p(Y, t)$ satisfies the following forward equation.

$$\begin{cases} \partial_t p = \frac{1}{2}\sum_{ij}\sigma_{ij}(Y)\partial_{ij}p - \sum_j \mu_j(Y)\partial_j p, \ Y \in \Omega \\ p(Y, t) = 0, \quad Y \in \partial\Omega \\ p(Y, 0) = \delta(X) \end{cases} \quad (9)$$

where $\partial_{ij}p = \frac{\partial^2 p(Y,t)}{\partial y_i \partial y_j}$, $\partial_i p = \frac{\partial p(Y,t)}{\partial y_i}$, $\partial_t p = \frac{\partial p(Y,t)}{\partial t}$. The $y_i$ is the i-th element of Y and $\delta(\cdot)$ is the two-dimensional dirac delta function. When m = 1, Eq. (8) becomes

$$C_p(\lambda, R) = \lambda V \pi R^2 \quad (10)$$

Now, given $\lambda$, we are interested in the joint optimization of X and R that minimize the total cost $C_t = C_u + C_p$, namely,

$$(X_{opt}, R_{opt}) = \arg\min_{0 \leq \|X\| < R < \infty}[C_u(X, R, \lambda) + C_p(\lambda, R)] \quad (11)$$

The difficulty in solving the joint optimization problem lies in obtaining the expression of $T(X, R, \lambda)$. Theorem 2 below indicates that $T(X, R, \lambda)$ is governed by a partial differential equation.

### A. Mean location update interval theorem

**Theorem 2:** (Mean location update interval theorem). Consider the MT's movement process $\{Y_t, t \geq 0\}$ stated in Theorem 1 with initial position $Y_0 = X \in \Omega$ and the call arrival interval $\zeta$. (a) Let $P(\zeta \geq t)$ denote the probability of $\zeta \geq t$. We have

$$T(X, R, \zeta) = \int_0^\infty G(X, t) P(\zeta \geq t) dt, \quad (12)$$

where $G(X, t) = P(\tau \geq t)$ is the probability that the MT with initial position X is still in $\Omega$ at time t. $G(X, t)$ is governed by the following backward equation.

$$\begin{cases} \partial_t G = \frac{1}{2}\sum_{ij}\sigma_{ij}(X)\partial_{ij}G + \sum_j \mu_j(X)\partial_j G, \ X \in \Omega \\ G(X, t) = 0, \quad X \in \partial\Omega \\ G(X, 0) = 1, X \in \Omega \\ \quad\quad\quad = 0, X \in \partial\Omega \end{cases} \quad (13)$$

where $\partial_{ij}G = \frac{\partial^2 G(X,t)}{\partial x_i \partial x_j}$, $\partial_i G = \frac{\partial G(X,t)}{\partial x_i}$, $\partial_t G = \frac{\partial G(X,t)}{\partial t}$.

(b) If $\zeta$ is an exponential random variable with parameter $\lambda$, then $T(X, R, \lambda)$ is governed by

$$\begin{cases} \frac{1}{2}\sum_{ij}\sigma_{ij}(X)\partial_{ij}T + \sum_j \mu_j(X)\partial_j T = -1 + \lambda T, \ X \in \Omega \\ T(X, R, \lambda) = 0, \quad X \in \partial\Omega \end{cases} \quad (14)$$

where $\partial_{ij}T = \partial^2 T(X, R, \lambda)/\partial x_i \partial x_j$, $\partial_i T = \partial T(X, R, \lambda)/\partial x_i$.

**Proof:** The main idea of the proof is that we regard $\{Y_t, t \geq 0\}$ as a diffusion process with an absorption boundary $\partial\Omega$ so that the equation of $T(X, R, \lambda)$ is connected to that of $\tau(X, R)$. Please refer to the APPENDIX for the complete proof.

**Corollary 1:** Under assumption A7, Eq. (14) becomes

$$\begin{cases} \frac{\sigma_{11}}{2}T_{xx} + \frac{\sigma_{22}}{2}T_{yy} + \mu_1 T_x = -1 + \lambda T, \ X \in \Omega \\ T(X, R) = 0, \quad X \in \partial\Omega \end{cases} \quad (15)$$

where $X = (x, y)$, $T_x = \frac{\partial T(X, R, \lambda)}{\partial x}$, $T_{xx} = \frac{\partial^2 T(X, R, \lambda)}{\partial x^2}$, and $T_{yy} = \frac{\partial^2 T(X, R, \lambda)}{\partial y^2}$. And we have

$$\mu_1 = \begin{cases} 0, & k \to 0 \\ \frac{E(\xi)}{E(\eta)}, & k \to \infty \end{cases} \quad (16)$$

$$\sigma_{22} = \begin{cases} \frac{E(\xi^2)}{2E(\eta)}, & k \to 0 \\ 0, & k \to \infty \end{cases}$$

$$\sigma_{11} = \begin{cases} \frac{E(\xi^2)}{2E(\eta)}, & k \to 0 \\ \frac{\text{Var}(\xi)E^2(\eta) + \text{Var}(\eta)E^2(\xi)}{E^3(\eta)}, & k \to \infty \end{cases}$$

**Proof:** We can easily obtain (15)-(16) substituting (2) into (1). Here, we omit the details.

**Remarks:** Previous work, such as [3], often neglects the location update due to call arrival. Theorem 2 explicitly embeds this update into a differential equation.

In general, to solve (9), (13)-(15), we can first transform them into a canonical form and then obtain its exact solution. However, the exact solution is often very complicated or difficult to calculate. For example, following the method in Ref [23], Eq. (15) can be transformed into a non-homogeneous modified Helmholtz equation with a homogeneous boundary condition over an elliptical domain. Although the exact solution to the Helmholtz equation over the elliptical domain can be expressed in terms of all eigenvalues and eigenfunctions of (15), only the first three eigenvalues and eigenfunctions can be easily approximately given (see Section 7.3.4-3 in Ref [23]). Also, only for some special case, the equation have a simple form. For example, when $k \to 0$ and $E(\xi^2) = 2E(\eta)$, $\{Y_t; t \geq 0\}$ becomes a standard Brownian motion. Further, when $\lambda = 0$, the exact solution $T(X, R, \lambda)$ to (15) is given by

$$T(X, R, \lambda) = \frac{R^2 - x^2 - y^2}{2}. \quad (17)$$

Fortunately, lots of numerical techniques (e.g., the finite element method [22]) can be used to solve these equations. However, we are more interested in the approximate closed-form solutions to these equations for the joint optimization because it can provide more direct and deep insights to the drift and the diffusion coefficient.

In this paper, due to space limit, we only focus on (7), (10),

and (15) for the joint optimization in (11). From (11), given R, the minimization problem of the total cost is transformed to the maximization problem of the mean location update interval over X. The basic idea to solve the joint optimization is the following. First, we obtain the approximate expression of T(X, R, λ). Second, the optimal X is expressed in terms of R. Finally, the joint optimization problem with two variables is reduced into a tractable optimization problem with one variable. Section IV.B below considers the joint optimization in (11) for the special case (i.e., $k \to 0$ and $k \to \infty$ when $\lambda = 0$) and Section IV.C below considers the joint optimization in (11) for the general case (i.e., $0 < k < \infty$ and $\lambda > 0$).

### B. Optimal X and R for the special cases

This section considers the optimal X and R for $k \to 0$ and $k \to \infty$ when $\lambda = 0$. Here, we set $\lambda = 0$ to approach the case of $\lambda \ll 1$ or $\zeta \gg \tau(X,R)$ (see (6)). Theorem 3 below gives the approximate expression of T (X, R, λ). Based on this expression, Theorem 4 below gives the optimal X and R and the minimum total cost; whereas Theorem 5 below gives the optimal R and the minimum total cost when X is fixed at the center of the LA.

**Theorem 3:** When $\lambda = 0$ in (15), T (X, R, λ) can be approximated by

$$T(X,R,\lambda) \approx \begin{cases} (R^2 - x^2 - y^2)(A + B(x+y)), & k \to 0 \\ C_1(y) + C_2(y)\exp(\dfrac{-2\mu_1 x}{\sigma_{11}}) - \dfrac{x}{\mu_1} + \dfrac{\sigma_{11}}{2\mu_1^2}, & k \to \infty \end{cases} \quad (18)$$

where $A = \dfrac{6(\sigma_{11}+\sigma_{22})}{(\mu_1 R)^2 + 6(\sigma_{11}+\sigma_{22})^2}$, $B = -\dfrac{\mu_1 A}{2(\sigma_{11}+\sigma_{22})}$,

$$C_2(y) = \dfrac{2\sqrt{R^2 - y^2}}{\mu_1[\exp(\dfrac{-2\mu_1\sqrt{R^2-y^2}}{\sigma_{11}}) - \exp(\dfrac{2\mu_1\sqrt{R^2-y^2}}{\sigma_{11}})]},$$

$$C_1(y) = -C_2(y)\exp(\dfrac{-2\mu_1\sqrt{R^2-y^2}}{\sigma_{11}}) + \dfrac{\sqrt{R^2-y^2}}{\mu_1} - \dfrac{\sigma_{11}}{2\mu_1^2}$$

**Proof:** Please refer to the APPENDIX.

The drift $\mu_1$ is "local", i.e., it is a drift in terms of each displacement. To consider the cumulative effect of the drift, we define a global drift to be $\gamma = 2\mu_1 R/\sigma_{11}$. From (16) and assumption A2, we have

$$\gamma = \begin{cases} 0, & k \to 0 \\ \infty, & k \to \infty \end{cases} \quad (19)$$

$\gamma$ plays a very important role in minimizing total cost and optimizing X and R. Note the initial position X is always at the x-axis in our coordinate system. We are now in the position to present the $X_{opt}(x_{opt}, 0)$ and $R_{opt}$ so as to minimize the total cost.

**Theorem 4:** When $\lambda = 0$ in (15), the $x_{opt}$, $T(x_{opt},R,\lambda)$,

$R_{opt}$, and the minimum total cost $C_t(X_{opt}, R_{opt}, \lambda)$ can be respectively approximated by

$$x_{opt} = \begin{cases} -\dfrac{3\gamma\sigma_{11}R}{8(\sigma_{11}+\sigma_{22})}, & k \to 0 \\ -R[1-\dfrac{\ln(2\gamma)}{\gamma}], & k \to \infty \end{cases} = \begin{cases} 0, & k \to 0 \\ -R, & k \to \infty \end{cases} \quad (20)$$

$$T(x_{opt},R,\lambda) \approx \begin{cases} R^2/(\sigma_{11}+\sigma_{22}), & k \to 0 \\ 2R/\mu_1, & k \to \infty \end{cases}$$

$$R_{opt} \approx \begin{cases} \sqrt[4]{(\sigma_{11}+\sigma_{22})U/(\lambda V\pi)}, & k \to 0 \\ \sqrt[3]{U\mu_1/(4\lambda V\pi)}, & k \to \infty \end{cases}$$

$$C_t(X_{opt},R_{opt},\lambda) \approx \begin{cases} 2\sqrt{(\sigma_{11}+\sigma_{22})\lambda UV\pi}, & k \to 0 \\ \sqrt[3]{(U\mu_1)^2\lambda V\pi/16}, & k \to \infty \end{cases}$$

**Proof:** Please refer to the APPENDIX.

**Remarks:** The conclusion on $X_{opt}$ in Theorem 4 is intuitive. If the movement is unbiased, an initial position at the center of the LA can maximize the mean location update interval. Whereas if the movement is a biased diffusion process with a positive preferred direction, the MT's initial position that is at the negative x-axis can maximize the mean update interval. In the extreme case, the initial position is near the boundary of the LA, (-R, 0). Also, from (20), when $k \to 0$, $\sigma_{11} \to 1$, and $\sigma_{22} \to 1$, $T(x_{opt},R,\lambda)$ approaches the exact solution in (17).

**Theorem 5:** When $\lambda = 0$ in (15), if X is fixed at the center of the LA, $R_{opt}$ and the minimum total cost $C_t(X_{opt}, R_{opt}, \lambda)$ can be approximated by

$$R_{opt} \approx \begin{cases} \sqrt[4]{(\sigma_{11}+\sigma_{22})U/(\lambda V\pi)}, & k \to 0 \\ \sqrt[3]{2}\sqrt[3]{U\mu_1/(4\lambda V\pi)}, & k \to \infty \end{cases} \quad (21)$$

$$C_t(X,R_{opt},\lambda) \approx \begin{cases} 2\sqrt{(\sigma_{11}+\sigma_{22})\lambda UV\pi}, & k \to 0 \\ \sqrt[3]{4}\sqrt[3]{(U\mu_1)^2\lambda V\pi/16}, & k \to \infty \end{cases}$$

**Proof:** Substituting X = (0,0) into (18) in Theorem 3, and following the proof of Theorem 4, we can obtain (21).

**Remarks:** Compare (20) for the case of optimal X and (21) for the case of fixed X at the center of LA. When $k \to 0$, $R_{opt}$ and $C_t(X_{opt}, R_{opt}, \lambda)$ of both cases are roughly equal. However, when $k \to \infty$, $R_{opt}$ in the latter is about $\sqrt[3]{2}$ ($\approx 1.260$) times that in the former, and $C_t(X_{opt}, R_{opt}, \lambda)$ in the later is about $\sqrt[3]{4}$ ($\approx 1.587$) times that in the former. It indicates that optimizing on initial position, which previous work did not consider, has the potential of reducing the cost measure by 37%.

### C. Optimal X and R for the general cases

This section considers the optimal X and R for the general case ($0 < k < \infty$ and $\lambda > 0$). We use the Galerkin method [22] to seek the approximate solution T (X, R, λ) to (15). The key of using the Galerkin method lies in how to construct the trial function. Clearly, the trial function under the general case should satisfy the boundary condition and the corresponding optimal initial position should have the properties in the limit case (see $x_{opt}$ in (20)). Let the trial function $g(x,y)$ be given by

$$g(x,y) = [\frac{R^2 - x^2 - y^2}{x+a}] \geq 0, \quad a = \frac{E(\xi)R}{E(\eta)\mu_1} \geq R \quad (22)$$

where $(x,y) \in \Omega$. We construct the first-order approximation solution T (X, R, λ) to (15) as follows.

$$T(X, R, \lambda) = C \cdot g(x,y) \geq 0 \quad (23)$$

where the constant coefficient C is given by

$$C \equiv \frac{-\iint_\Omega dxdy}{\iint_\Omega Lg dxdy} = \frac{-\pi R^2}{\frac{\sigma_{11}}{2}C_{11} + \frac{\sigma_{22}}{2}C_{22} - \lambda C_0} \geq 0 \quad (24)$$

$$Lg = \frac{\sigma_{11}}{2}g_{xx} + \frac{\sigma_{22}}{2}g_{yy} + \mu_1 g_x - \lambda g$$

$$C_{11} = -2a\int_{-R}^{R} \frac{2\sqrt{R^2 - y^2}}{a^2 - (R^2 - y^2)} dy$$

$$C_{22} = -2\int_{-R}^{R} \ln \frac{a + \sqrt{R^2 - y^2}}{a - \sqrt{R^2 - y^2}} dy$$

$$C_0 = \int_{-R}^{R} (R^2 - y^2 - a^2) \ln \frac{a + \sqrt{R^2 - y^2}}{a - \sqrt{R^2 - y^2}} dy + a\pi R^2$$

Taking the first-order and second-order derivations of T (X, R, λ) with respect to x in (23) and substituting $X = (x, 0) \in \Omega$ into them, we have

$$T_x = Cg_x(x,0) = C\frac{-(x^2 + 2ax + R^2)}{(x+a)^2} \quad (25)$$

$$T_{xx} = Cg_{xx}(x,0) = C\frac{2(R^2 - a^2)}{(x+a)^3} \leq 0 \quad (26)$$

Eq. (26) indicates that the solution x to equation $T_x = 0$ can maximize T (X, R, λ). Noting that

$$a = \frac{E(\xi)R}{E(\eta)\mu_1}R = \begin{cases} R, k \to \infty \\ \infty, k \to 0 \end{cases} \quad (27)$$

and solving $T_x = 0$, the optimal x, $x_{opt}$, that is within (-R, R) is given by

$$x_{opt} = -a + \sqrt{a^2 - R^2} \quad (28)$$

$$= -\frac{R^2}{a + \sqrt{a^2 - R^2}}$$

$$= \begin{cases} 0, \ k \to 0 \\ -R, k \to \infty \end{cases}$$

Clearly, we also have $x_{opt} \in (-R, 0]$.

Now, the optimal initial position is already expressed in terms of R. Then the joint optimization problem with two variables is reduced into a tractable optimization problem with one variable. Further, we can use many mature methods to find the optimal R and the corresponding minimum total cost.

D.     *Model extension and further explanations*

With our analytical framework, we can evaluate the location management cost through solving the partial differential equation (see (9) (13) (14)) which can always be done by many mature numerical techniques [22]. Here, we emphasize the following advantages of our methods.

(I) Our method can easily deal with the general call arrival distribution (see (13)), the general displacement length distribution, the general displacement time distribution, and the general displacement direction distribution (see (1)).

(II) Our method can easily deal with the LA with different shape which is embedded in the boundary condition of the partial differential equation. The reason is that both numerical methods (e.g., the finite element method [22]) and analytically approximate methods (e.g., the Galerkin method [22])) of solving the partial differential equation only require that the LA be a simply connected region with continuous boundary. Therefore, our analytical framework does not impose special restrictions on the shape of the LA. For example, in the Galerkin method, if the shape of the LA can be expressed as g (x, y) = 0, then the approximate solution can be expressed as z (x, y) g (x, y), where z (x,y) is to be determined (see the proof in Theorem 3).

(III) Theorem 2 is true if MT's movement can be model as an n-dimensional homogeneous diffusion process.

(IV) Our method is more general and more powerful compared with the Markov chain method. The CTRW model generalizes the classical "discrete" random walk model. Therefore, our method based on the CTRW model should produce more general result than the Markov chain method. In the following, we will restore a result which has been obtained in [3] using Markov chain method.

Consider the movement of an MT in the one-dimensional interval $\Omega = [0, L]$. Assume the initial position of the MT is $x \in \Omega$. With a preferred direction toward to the end point L, the MT travels along a road section of a random length $\xi$. At the end of the road section, it selects a forward direction with probability p, or a backward direction with probability q, where $p + q = 1$. The displacement time for each road section is i.i.d. with $\eta$. When the MT arrives one of end points 0 and L, it will performs a location update. Other assumptions follow the A2-A6. Now, we define

$$\xi' = \begin{cases} \xi, & \text{with probablity p} \\ -\xi, & \text{with probablity q} \end{cases} \quad (29)$$

Using Theorem 1, we have

$$\mu = \frac{E(\xi')}{E(\eta)} \text{ and } \sigma = \frac{Var(\xi')E^2(\eta) + Var(\eta)E(\xi')^2}{E^3(\eta)}. \quad (30)$$

When the call arrival interval is the Poisson distribution with parameter λ, using the conclusion (b) in Theorem 2, we have

$$\begin{cases} (\sigma/2)T_{xx} + \mu T_x = -1 + \lambda T, \, x \in \Omega \\ T(0,L) = T(L,L) = 0 \end{cases} \quad (31)$$

We can easily find the exact solution of (31) for any λ. Due to space limit, we only present the exact T (x, L) when λ = 0 for restoring a result in [3]. When λ = 0, T(x, L) governed by (31) is given by

$$T(x,L) = \frac{L(1-\exp(\frac{-2\mu x}{\sigma})) - x(1-\exp(\frac{-2\mu L}{\sigma}))}{\mu(1-\exp(\frac{-2\mu L}{\sigma}))} \quad (32)$$

From (31), we can easily obtain the optimal initial position $x_{opt}$ and the corresponding T ($x_{opt}$, L) as follows

$$x_{opt} = -\frac{\sigma}{2\mu}\ln[\frac{1-\exp(-2\mu L/\sigma)}{2\mu L/\sigma}] \to L/2, \quad \mu \to 0 \quad (33)$$

$$T(x_{opt}, L) \to L^2/(4\sigma), \quad \mu \to 0$$

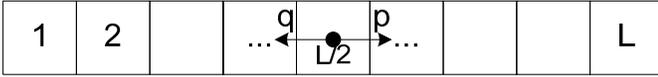

Fig. 4. The Markov chain model for the one-dimensional case.

Now we consider the classical discrete random walk model for the one-dimensional case (see Fig. 4). In this model, the coverage of cellular networks is partitioned into many cells of equal length and an MT moves into one of adjacent cells with probability p and q at each discrete time slot t, respectively. The numbers in Fig. 5 represent the numbering of cells. The location update area is the area including cell 1 to cell L. With the initial position L/2, when the MT moves to one of adjacent cells with equal probability, i.e., p = q = 1/2, without considering the location update cost cased by call arrival, i.e., λ = 0, using Markov chain method with the state space {1,2,…, L}, the mean location update time obtained in [3] is equal to $L^2/4$.

In contrast, under the same assumption, substituting p = q = 1/2, E (ξ) = 1, E (η) = 1 into (30) and (33), we have σ = 1 and T (L/2, L) = $L^2$/4.

(V) We now discuss assumption A2, i.e., how large is the radius R so that Theorem 1 holds.

Let n (t) denote the number of the displacements at time t, where n (t) is a random variable. Let E (n (t)) denote the mean of n (t). From Ref. [9], E (n (t)) can be approximately calculated by t/E (η), where E (η) is the mean of each displacement time.

The key condition that Theorem 1 holds is a sufficiently large n (t) rather than $R \gg E(\xi)$. Note that $R \gg E(\xi)$ necessarily leads to a large n(t), where $t = T(x_{opt}, R_{opt}, \lambda)$. $T(x_{opt}, R_{opt}, \lambda)$ is the mean location update interval that an MT updates its LA defined by the optimal initial position and the optimal radius. We now estimate E (n (t)), where $t = T(x_{opt}, R_{opt}, \lambda)$. We focus on the case of $k \to 0$ since other cases are similar. Noting $\sigma_{11}$ & $\sigma_{22}$ in (16) and $T(x_{opt}, R, \lambda)$ & $R_{opt}$ in (20), and using the default parameter values in Table 1, we have

$$E(n(T(x_{opt}, R_{opt}, \lambda))) \approx \frac{T(x_{opt}, R_{opt}, \lambda)}{E(\eta)}$$

$$= \frac{R_{opt}^2}{\sigma_{11} + \sigma_{22}} \frac{1}{E(\eta)}$$

$$= \sqrt{\frac{U}{\pi V}} \frac{1}{\sqrt{\lambda E(\xi^2) E(\eta)}}$$

$$\approx 1338$$

$$R_{opt} \approx \sqrt[4]{\frac{U}{\pi V} \cdot \frac{\sigma_{11} + \sigma_{22}}{\lambda}}$$

$$= \sqrt[4]{\frac{U}{\pi V} \cdot \frac{E(\xi^2)}{\lambda E(\eta)}}$$

$$\approx 1.03 \text{ km}$$

Therefore, under the weak drift, when the mean length of each road section is equal to 20 m, the optimal radius of the LA is about equal to 1.03 km. However, when the MT moves out of the optimal LA, the mean number of the displacements approaches 1338 that is sufficiently large. In other words, a moderate size of R (rather than $R \gg E(\xi)$) is enough to guarantee that Theorem 1 holds.

## V. NUMERICAL RESULTS

In this section, we examine the numerical results of (7), (10), (11), (15), and (23). We assume that the call arrival interval is an exponential distribution with parameter λ and the displacement length is also an exponential distribution with parameter h. The $f_\Theta (k, \theta)$ is given in (3). We use the following parameters: λ (unit: calls per hour), R (unit: km), 1/h (unit: m), E (η) (unit: s), Var (η) (unit: s), U, and V. The default parameter values are shown in Table 1.

Table 1: The default parameter values

| λ | R | 1/h | E (η) | Var (η) | U | V |
|---|---|-----|-------|---------|---|---|
| 2 | 1 | 20  | 8     | 1       | 20| 1 |

As explained in Section IV.A, we do not plot the exact solution to (15). In the following, Figs. 5 and 6 explain the relationship between the mean location update interval and the call arrival rate when the drift varies (i.e., when k is different). Figs. 7 and 8 show the optimal initial position, the optimal radius, and the cost saving ratio under the joint optimization, where the cost saving ratio is defined as follows.

$$\text{Cost saving ratio} = \frac{C_t(X_{center}, R_{opt}, \lambda) - C_t(X_{opt}, R_{opt}, \lambda)}{C_t(X_{center}, R_{opt}, \lambda)}. \quad (34)$$

where $C_t(X_{center}, R_{opt}, \lambda)$ denotes the minimal total cost under the optimal radius $R_{opt}$ when the initial position $X_{center}$ is always the center of the LA. In contrast, $C_t(X_{opt}, R_{opt}, \lambda)$ denotes the minimal total cost under the optimal initial position $X_{opt}$ and the optimal radius $R_{opt}$. They can be calculated based on (7), (10), (11), (23), and (28).

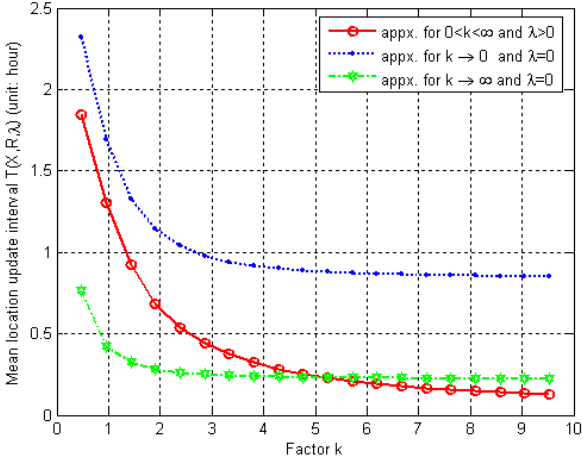

Fig. 5. The mean location update interval vs. factor k for different approximations.

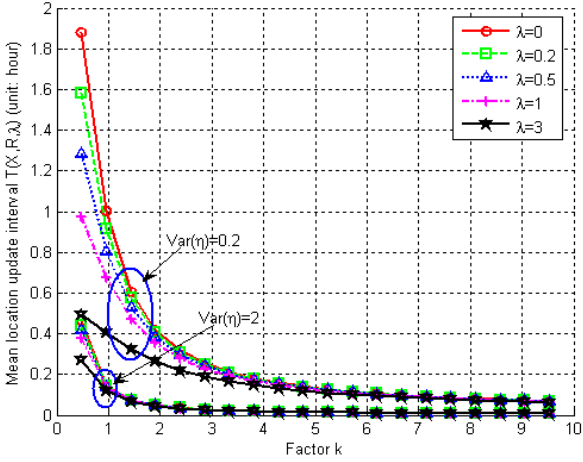

Fig. 6. The mean location update interval vs. factor k for different call arrival rates.

Fig. 5 plots (23) and the second equation in (20) when factor k varies, where Var($\eta$) = 0.1 s. Because the second equation in (20) is obtained when $\lambda \ll 1$, we set $\lambda$ = 0.2 calls per hour. From this figure, the results when k $\rightarrow$0 and k $\rightarrow \infty$ can well approach the general result in (23).

Fig. 6 plots (23) when $\lambda$ = 0, 0.2, 0.5, 1, 3 calls per hour and Var ($\eta$) = 0.2 s or 2 s. Note that $\lambda$ = 0 corresponds to the case that we ignore the location update incurred by call arrival. For the case of weak drift (i.e., k $\rightarrow$0), from this figure, the mean location update interval shows distinct differences when Var ($\eta$) = 0.2s and Var ($\eta$) = 2s, which indicates the mean interval is very sensitive to Var ($\eta$). This can be explained as follows: From (1), given E ($\eta$), the larger the Var ($\eta$), the higher the diffusion speed, and therefore the shorter the exit time. When Var ($\eta$) is smaller, the MT will spend more time in its LA so that the call arrival interval will play a more role in the mean location update interval (see Fig. (6)). Therefore, under this case, ignoring call arrival interval (i.e., $\lambda$ = 0) will cause a bigger error. For the case of strong drift (i.e., k $\rightarrow \infty$), from this figure, the curves under different $\lambda$ almost overlap when Var ($\eta$) = 0.2 s or 2 s, which indicates that ignoring call arrival interval (i.e., $\lambda$ = 0) will not cause a bigger error. The reason is that under this case, the MT can quickly exit from the LA so that the call arrival interval will play a less role in the mean location update interval (i.e., $\zeta \gg \tau(X,R)$, see Fig. (6)).

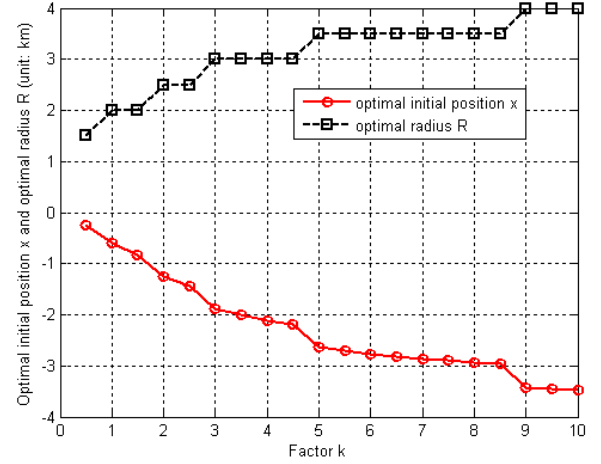

Fig. 7. The optimal initial position and the optimal radius vs. factor k.

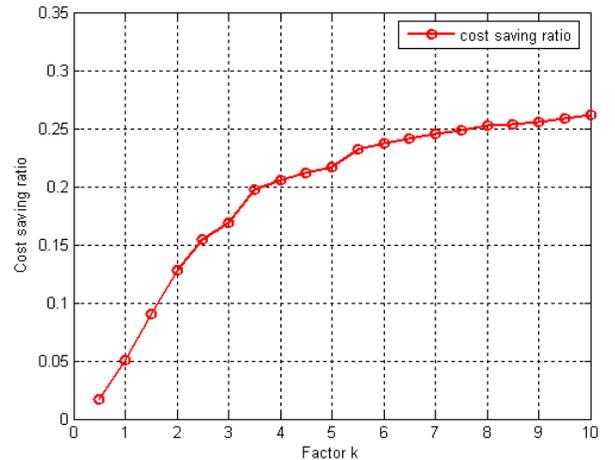

Fig. 8. The cost saving ratio vs. factor k.

Fig. 7 plots the optimal initial position achieved in (28) and the optimal radius R that minimizes the total cost. Fig. 8 plots the cost saving ratio defined in (34). Figs. 7 and 8 are illustrated based on the default parameter values. From Fig. 7, as explained previously, the optimal initial position approaches the origin of the LA for the case of weak drift whereas the optimal initial position approaches the boundary of the LA (i.e., (-R, 0)) for the case of strong drift. From Fig. 8, the cost saving ratio increases when the drift becomes stronger. As explained in Section IV.B, this ratio will be bounded by 37%. Also, the cost saving ratio can achieve 25% for a moderate drift. Therefore, it is worthwhile optimizing the initial position.

## VI. IMPLEMENTATION ISSUES

We have so far considered the analysis of the well-known distance-based location management scheme under the CTRW mobility model. We have proved a general result that using the optimal initial position has a potential of reducing the total cost by 37% (see remarks for Theorem 5). And the numerical results have shown a reduction of 25% for a moderate drift (see Fig. 8). The theoretical results have also indicated that our design of LA is always better than or equal to the usual design which only optimize the distance threshold. This section is devoted to a brief discussion of implementation issues: that is, how to integrate the theoretical results into an engineering system.

In our system architecture, the call arrival interval is assumed to be an exponential distribution with parameter $\lambda$. The k in (3) is chose based on the propensity of the preferred direction. For example, k = 0.5 corresponds to a weak drift and k=20 corresponds to a strong drift. Further, let $Y_\tau$ be the initial position (the location that the MT performs its last location update) with respect to the new LA. Let D denote the preferred direction which is always kept by the MT during its movement. The other assumptions and notations are shown in Section II.

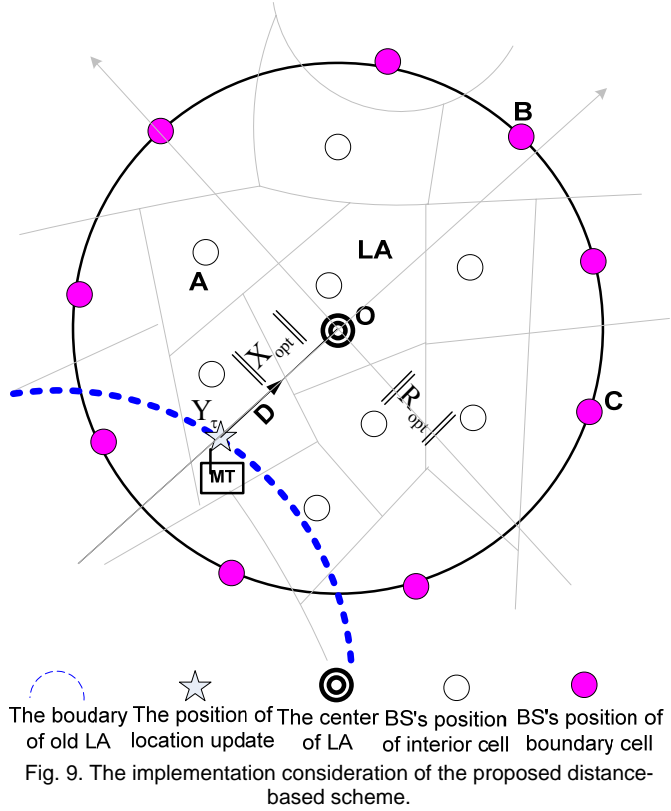

Fig. 9. The implementation consideration of the proposed distance-based scheme.

The system inputs are three sets of physical parameters (PM): PM1={k, E ($\xi$), Var ($\xi$), E ($\eta$), Var ($\eta$)}, PM2={$Y_\tau$, D}, and PM3={$\lambda$, U, V}. Like [5] [12], we assume that the MT can estimate these parameters. The reader can also refer to [20] [21] for the parameter estimation method. With PM1 and PM3, the ($X_{opt}$, $R_{opt}$) can be calculated based on (7), (10), (11), (23), and (28). Further, the LA can be formed with ($X_{opt}$, $R_{opt}$) and PM2.

The system outputs are two sets of cell IDs defining the LA of MT, consisting of "interior cells" and "boundary cells". The set of interior cells is composed of m sets of "sub-area cells", where m is the maximum paging delays. These sets will be defined below. Based on the outputs, the system performs location update operation and terminal paging operation with delay constraint.

We now consider **the construction of the new LA** (see Fig. 9). More details will be discussed later. Let the new x-axis pass through the point $Y_\tau$ and let the positive direction of the x-axis is the preferred direction D. Let the point, whose positive offset from point $Y_\tau$ equals to $\|X_{opt}\|$, be the new origin O. Then the new origin O and the radius $R_{opt}$ define the new LA. In the new LA, $Y_\tau$ becomes the new initial position, and $R_{opt}$ becomes the new distance threshold. From this construction, the center O of the new LA is not always the MT's new initial position. In contrast, in usual design of the LA, the MT's initial position $Y_\tau$ is always the center O of the new LA. Then the $Y_\tau$ and the radius $R_{opt}$ define the new LA.

After the new LA is formed, it is still difficult to use the LA for location management. The reason is that the formed LA is a connected region (the LA is circular disc here) with a continuous boundary. However, in reality, the coverage of PCS networks consists of many discrete cells. Therefore, we need to discretize the LA into a set of cells.

We now define **the boundary cell** and **the interior cell** of LA. We call a cell a boundary cell of an LA if the distance between the coordinate of the cell's base station and the coordinate of the LA's center O exceeds the distance threshold $R_{opt}$. Otherwise, we call the cell an interior cell. For example, in Fig. 9, cell A is a interior cell, and cell B and cell C are the boundary cells. An MT is always associated with a cell through the cell's base station. So, the MT can always find out the cell which it is currently residing at easily. As the MT moves and is handed off from cell to cell, it maintains the ID of its associated cell. Each MT stores its LA's boundary cell IDs in its memory. By the ID of the boundary cells, the MT can judge whether it has moved out of its current LA. The network stores the interior cell IDs of each MT's LA in the location database. By the ID of the MT (or the index of the entry in the location database), the network can update the LA of the MT and page the MT in the LA of the MT.

We proceed to define **the sub-area cells** for the paging scheme with delay constraint. Assume that the maximum paging delays is m. The sub-paging areas PA1, PA2, …, and PAm have been partitioned using the method in Section III.B. We call an interior cell A the sub-area cell of a sub-paging area PAi if the coordinates of cell A is within PAi. All such cells of PAi form a sub-area-cell_list of PAi. Then each sub-paging area corresponds to a unique sub-area-cell_list. And the m sets of "sub-area cells" constitute the set of interior cells.

**Remarks:** Since the network has the coordinates of all cells' base station, it can then determine easily the boundarycell_list, the interiorcell_list, and the sub-area-cell_list. Although the discretization method on the LA causes some inaccuracy, it greatly simplifies the difficulty of the implementation of the distance-based scheme.

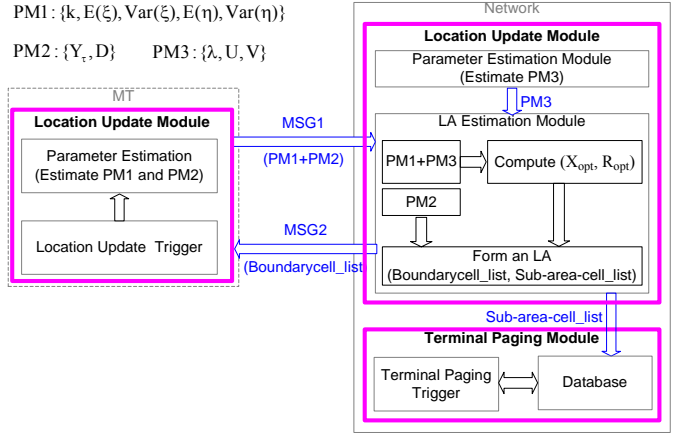

Fig. 10. The implementation Modules of our proposed distance-based scheme.

We are now in the position to present the system architecture (see Fig. 10). The MT, in collaboration with the network, performs the location update procedure. The location update module in the MT consists of "Location Update Trigger" sub-module and "Parameter Estimation" sub-module. The location update module in the network consists of the "LA Estimation" sub-module and "Parameter Estimation" sub-module. In addition, the terminal paging module in the network is responsible for sending a paging message to identify the current cell location of the MT. We outline them as follows.

**The location update procedure:** The location update can be triggered by the MT or the network. For a focus, we describe the MT-triggered implementation below; the main concept also applies to the network-triggered implementation – it is all a matter of who is responsible for keeping track of the state of the MT.

The MT operates as follows. When entering a cell, the MT will receive a cell ID broadcasted by the cell's base station. If the received cell ID is in the boundary cell list, the location update procedure will be triggered by the MT's "Location Update Trigger" sub-module. Under this condition, the sub-module will immediately call the "Parameter Estimation" sub-module to estimate the PM1 and PM2. After assembling the PM1 and PM2 into a piece of message (see MSG1 in Fig. 10), the MT then sends this message to the network. If the received cell ID is not in the boundary cell list, the MT only computes some necessary information depending on the parameter estimation method.

The network operates as follows. Upon receiving the LU message from the MT, the network first estimates the PM3 by calling its "Parameter Estimation" sub-module. From the PM1 and PM3, the network next computes $X_{opt}$ and $R_{opt}$. By the $X_{opt}$, $R_{opt}$, and the PM2, the network now can form a new LA and further produce the boundarycell_list, the interiorcell_list, and the sub-area-cell_list.. Finally, the network assembles the boundary cell list into a piece of message (see MSG2 in Fig. 10), then sends this message to the MT. When the MT receives the boundary cell list, it stores the list in its memory so that it knows when to trigger the next location update in the future.

**The terminal paging procedure:** When a call to identify the location of an MT arrives, with the sub-area-cell_list, the network can sequentially page the sub-paging area to identify the location of the MT.

## VII. CONCLUSIONS

This paper addresses the problem of the optimal design of LAs to reduce location management cost. With the popular CTRW mobility model, we propose a novel framework to minimize the total location management cost. In this framework, the optimization problem of location management is transformed into a problem of solving partial differential equation. We optimize on two physical attributes of the LA: 1) the size of the LA; and 2) the initial-position offset of an MT from the center of the new LA as it crosses the boundary of the old LA. The latter is neglected in most previous research, and we show that this is a design parameter that should not be ignored. Specifically, when the LA is a circular disc and the Poisson call arrival rate is sufficiently small, we prove a general result that when the drift is strong (there is a strong directional preference in the MT movement), using the optimal offset can achieve as much as a reduction of 37% on the overall location management cost, regardless of the relative magnitudes of the underlying two constituent costs, terminal paging and LA update costs.

## VIII. ACKNOWLEDGEMENTS

The authors would like to thank the anonymous reviewers for their helpful suggestions and insightful comments on this paper.

APPENDIX: A PROOF OF THEOREM 2

Note that $\zeta$ is independent of $\tau$. We have

$$T(X,R,\lambda) = E(\min(\tau(X,R),\zeta)) \qquad (35)$$

$$= \int_0^\infty P(\min(\tau,\zeta) \geq t)dt$$

$$= \int_0^\infty P(\tau \geq t, \zeta \geq t)dt$$

$$= \int_0^\infty P(\tau \geq t)P(\zeta \geq t)dt$$

By Theorem 1, the original movement $\{(X_i, T_i), i \geq 1\}$ of the MT can be approached by a diffusion process $\{Y_t; t \geq 0\}$. Further, in the distance-based scheme, the original movement can be regarded as a diffusion process with an "absorbing boundary". The reason is the following. In the distance-based scheme, if an MT crosses the boundary $\partial\Omega$, a new LA will be formed, which means that the MT will disappear from its old LA. In other words, the MT will be absorbed by the boundary $\partial\Omega$. Now, we let $\{Y_t; t \geq 0\}$ denote the diffusion process with the absorbing boundary $\partial\Omega$, where $Y_0 = X \in \Omega$. Let G (X, t) be the probability that the MT with initial position X is still in $\Omega$ at time t. G(X, t) is also equal to the probability that the exit time $\tau$ (X, R) is equal to or larger than t, i.e., G(X, t) = P($\tau \geq$ t). By the first exit theory [24], G(X, t) is governed by the backward equation in (13). When $\zeta$ is an exponential distribution with parameter $\lambda$, we have

$$T(X,R,\lambda) = \int_0^\infty e^{-\lambda t} G(X,t)dt \qquad (36)$$

Note that

$$\int_0^\infty e^{-\lambda t} \frac{\partial G(X,t)}{\partial t} dt = -1 + \lambda T . \qquad (37)$$

Multiplying both sides of the first equation in (13) by $e^{-\lambda t}$ and integrating (13) over $(0,\infty)$ with respect to t, we have

$$\frac{1}{2}\sum_{ij}\sigma_{ij}(X)\partial_{ij}T + \sum_j u_j(X)\partial_j T = -1 + \lambda T, X \in \Re \qquad (38)$$

Eq. (14) is obtained from (38) and the boundary condition in (13). □

APPENDIX: A PROOF OF THEOREM 3

If $k \to 0$, we use the two-order Galerkin method [22] to solve (15). Noting the boundary condition, we approach the solution to (15) by

$$T(X,R,\lambda) = A\varphi_1 + B\varphi_2 \qquad (39)$$

$$\varphi_1 = R^2 - x^2 - y^2, \varphi_2 = (R^2 - x^2 - y^2)(x+y).$$

where $\varphi_i$ is called the trial function and the constants A and B satisfy the following equation

$$\begin{pmatrix} \iint_\Omega \varphi_1 L\varphi_1 & \iint_\Omega \varphi_1 L\varphi_2 \\ \iint_\Omega \varphi_2 L\varphi_1 & \iint_\Omega \varphi_2 L\varphi_2 \end{pmatrix} \begin{pmatrix} A \\ B \end{pmatrix} = \begin{pmatrix} -\iint_\Omega \varphi_1 \\ -\iint_\Omega \varphi_2 \end{pmatrix}. \qquad (40)$$

where $L(\cdot) = [\sigma_{11}/2](\cdot)_{xx} + [\sigma_{22}/2](\cdot)_{yy} + \mu_1(\cdot)_x$.

Solving (40), we have

$$A = \frac{6(\sigma_{11}+\sigma_{22})}{(\mu_1 R)^2 + 6(\sigma_{11}+\sigma_{22})^2} \quad B = -\frac{\mu_1 A}{2(\sigma_{11}+\sigma_{22})} \qquad (41)$$

If $k \to \infty$, we have $\lim_{k\to\infty}\sigma_{22} = 0$. Using the regular perturbation method [22], we approach the solution to (14) by

$$T = \bar{T} + o(\sigma_{22}). \tag{42}$$

where $\bar{T}(X, R, \lambda)$ is governed by (43)

$$\begin{cases} [\sigma_{11}/2]\bar{T}_{xx} + \mu_1 \bar{T}_x + 1 = 0, & \|X\| < R \\ \bar{T}(X, R, \lambda) = 0, & \|X\| = R \end{cases} \tag{43}$$

Solving (43), we easily obtain the result when $k \to \infty$ in (18). □

## APPENDIX: A PROOF OF THEOREM 4

Noting the initial position of the MT X is always at the x-axis, we substitute $(x, 0)$ into (18). We easily know $T_{xx}(X, R, \lambda) < 0$, which means that the solution that satisfies $T_x(X, R, \lambda) = 0$ can maximize the $T(X, R, \lambda)$. Let $x_{opt}$ denote the solution. Then $x_{opt}$ is given by

$$x_{opt} = \begin{cases} \dfrac{R(1 - \sqrt{1 + 3[\dfrac{\gamma \sigma_{11}}{4(\sigma_{11} + \sigma_{22})}]^2})}{\dfrac{3\gamma \sigma_{11}}{4(\sigma_{11} + \sigma_{22})}}, & k \to 0 \\ \dfrac{-R}{\gamma} \ln[\dfrac{\exp(\gamma) - \exp(-\gamma)}{2\gamma}], & k \to \infty \end{cases} \tag{44}$$

$$\sim \begin{cases} -\dfrac{3\gamma \sigma_{11} R}{8(\sigma_{11} + \sigma_{22})}, & k \to 0 \\ -R[1 - \dfrac{\ln(2\gamma)}{\gamma}], & k \to \infty \end{cases}$$

$$\to \begin{cases} 0, & k \to 0 \\ -R, & k \to \infty \end{cases}$$

where "~" denotes "same order".

Substituting the first equation in (44) into (18), we have

$$T(x_{opt}, R, \lambda) \to \begin{cases} \dfrac{R^2[48(\sigma_{11} + \sigma_{22})^2 + 9(\gamma \sigma_{11})^2]}{48(\sigma_{11} + \sigma_{22})^3}, & k \to 0 \\ 2R/\mu_1, & k \to \infty \end{cases} \tag{45}$$

$$\to \begin{cases} R^2/(\sigma_{11} + \sigma_{22}), & k \to 0 \\ 2R/\mu_1, & k \to \infty \end{cases}$$

Substituting the last first equation in (45) into (7), we have
$$C_t(x_{opt}, R, \lambda) = U/T(x_{opt}, R, \lambda) + \lambda \pi R^2 V \tag{46}$$

$$\to \begin{cases} \dfrac{U(\sigma_{11} + \sigma_{22})}{R^2} + \lambda \pi R^2 V, & k \to 0 \\ \dfrac{U\mu_1}{2R} + \lambda \pi R^2 V, & k \to \infty \end{cases}$$

$$\geq \begin{cases} 2\sqrt{(\sigma_{11} + \sigma_{22})\lambda UV\pi}, & k \to 0 \\ \sqrt[3]{(U\mu_1)^2 \lambda V \pi / 16}, & k \to \infty \end{cases}$$

Note the condition that the equality of the last equation in (46) holds. We have

$$R_{opt} \approx \begin{cases} \sqrt[4]{(\sigma_{11} + \sigma_{22})U/(\lambda V \pi)}, & k \to 0 \\ \sqrt[3]{U\mu_1/(4\lambda V \pi)}, & k \to \infty \end{cases} \tag{47}$$